\documentclass{article}
\usepackage{sao2,psfig}

\def\lea{\ifmmode ^{<}_{\sim} \else $^{^{<}_{\sim}}$\fi}
\def\gea{\ifmmode ^{>}_{\sim} \else $^{^{>}_{\sim}}$\fi}
\def\sec{\ifmmode {}^{\prime\prime}\else ${}^{\prime\prime}$\fi~}

\def\magdot{\ifmmode {}^{\rm m}\!\!\!.\, \else ${}^{\rm m}\!\!\!.\,$\fi}
\def\asec{\ifmmode ^{\prime\prime}\else$^{\prime\prime}$\fi}
\begin{document}

\title{Magnetic field measurements in white dwarfs. Magnetic field,
    rotation and spectrum of 40 Eri B}
\author{S.N. Fabrika, G.G. Valyavin, T.E. Burlakova,
E.A. Barsukova, D.N. Monin }
\institute {\saoname}
\maketitle
\begin{abstract}
This paper describes results of magnetic field measurements of white dwarfs
carried out on the 6--m telescope for the last years. A magnetic field
of about $B_e \approx
28$~kG has been discovered in the degenerate star WD\,1953--011 which
has been selected from high--resolution spectroscopy of non--LTE H$\alpha$
core by Koester et al. (1998). A rotational period of WD\,0009+501, 1.83 hours,
has been discovered, the average magnetic field of the star is
$<B_e> = -42.3 \pm 5.4$~kG and its semi--amplitude of the rotational
variability is $32.0 \pm 6.8$~kG. The variable magnetic field of the
bright normal (non--magnetic) degenerate star 40~Eridani~B was
confirmed in January 1999 by Zeeman time--resolved spectroscopy.
Both the H$\alpha$ and the H$\beta$ lines give about the same results,
we have selected two best periods in the magnetic field
variability, ${\rm 2^h\,25^m}$ and ${\rm 5^h\,17^m}$.
The  semi--amplitude of the rotational variations $B_{max} \approx 4000
\div 5000$~G and the average field is about zero $\pm 500$~G.
If the magnetic field of 40~Eridani~B is a central dipole, then the rotational
axis inclination to the line of sight is $i \sim 90^{\circ}$, and
the magnetic axis inclination to the rotational axis is about the same,
$\beta \sim 90^{\circ}$. For the first time an ultra--high signal--to--noise
spectrum of the white dwarf has been obtained (S/N~$ >$~1000). We have
found in this hydrogen--rich DA white dwarf 40~Eridani~B (16500~K) that
helium abundance is low (${\rm N(He)/N(H) < 10^{-7}}$), but the
 spectrum is rich in ultra--weak
absorption lines of metals in low ionization states. It was proposed
that these lines were produced in both circumstellar and interstellar
gas. The gas may be supplied by  accretion from interstellar medium
and from the dM4e companion 40~Eridani~C with an accretion rate $\dot{M_a}
\sim  10^{-19}\, M_{\sun}$/y. The accreting gas may
form a circumstellar rotating envelope in the magnetosphere at a distance of
$\sim 4\,\cdot 10^{11}$~cm.

\end{abstract}
\section*{Introduction}

The magnetic observations of dwarfs (WDs) have been carried out  in the
Laboratory of Stellar Physics of SAO since 1989. Originally the tasks were
to detect magnetic fields in normal DA WDs observing them
with an accuracy of about a few kG and to study the distribution of WDs
over their surface magnetic fields (magnetic field function),  to observe
known magnetic WDs in order to find their rotational periods and
to study the magnetic fields~-- rotation relation. Until 1995 the
observations were fulfilled with the hydrogen--line magnetometer of
the 6--m telescope (Stol', 1991; Bychkov et al., 1991; Shtol' et al.,
1997; Fabrika et al., 1997). These observations and such observations of
other authors (mainly by Schmidt \& Smith, 1995) have shown that
WDs with magnetic fields $B \ga 10$~kG are not numerous.

It was found
in the later statistical studies (Fabrika \& Valyavin, 1999; Valyavin
\& Fabrika, 1999) that the frequency of magnetic $B \ga 1$~MG WDs is
about 2~$\%$ among hot (young stars) and it is about 20~$\%$ among
cool (old) stars, i.~e. the magnetic field does certainly evolve in WDs.
It was also found that the magnetic field function is a power one,
$P_B \sim B^{-\alpha}$, with a power index $\alpha \approx 1.3$.
Normalization properties of the magnetic field function allowed to
estimate the frequency of magnetic WDs in the weak--field limit. We
may expect the minimum {\it surface} magnetic field strength to be 1--10~kG
in hot WDs ($T>10000$~K) and in cool WDs ($T<10000$~K) it is 10--50~kG.
To confirm
these conclusions by direct observations and to study the weak--field
part of the magnetic field function, we decided to observe the brightest DA
WDs with a higher accuracy.

\section*{Zeeman spectroscopy with the Main Stellar Spectrograph}

In 1995--1997 Zeeman spectral observations of bright DA WDs were
carried out on the Main Stellar Spectrograph (MSS) of the 6--m telescope
with a circular polarization analyzer (Valyavin et al., 1997;
Fabrika \& Valyavin, 1999). The time--resolved spectroscopy
with 3~-- 5~min exposure times was aimed at both detection of possible
rotational variability in individual spectra and deriving
high accuracy estimates of average magnetic field in sum spectra.
Each image in this mode of observations consists of two simultaneous
spectra of opposite circular polarization which are splitted in the analyzer.
Two different orientations of the first quarter--wave plate are used to record
opposite circular polarization spectra at the same place of the
CCD detector in two adjacent exposures.
We used
this configuration of the higher spectral resolution with a hope to
detect magnetic shifts in narrow non--LTE H$\alpha$ profiles
of DA ($8000 \div 22000$~K) WDs. It was expected that observations in this
mode could provide the desired accuracy of about 1~kG. In some brightest
WDs such an accuracy has actually been  reached, however without
positive detections.
For example in time--resolved Zeeman spectroscopy of WD\,0713+584 we did
not detect any periodical signals during a time of observations of 113~min.
The total average effective magnetic field estimate has been obtained
in this star particular $B_e = 0.1 \pm 1.0$~kG.

\begin{figure}
\psfig{figure=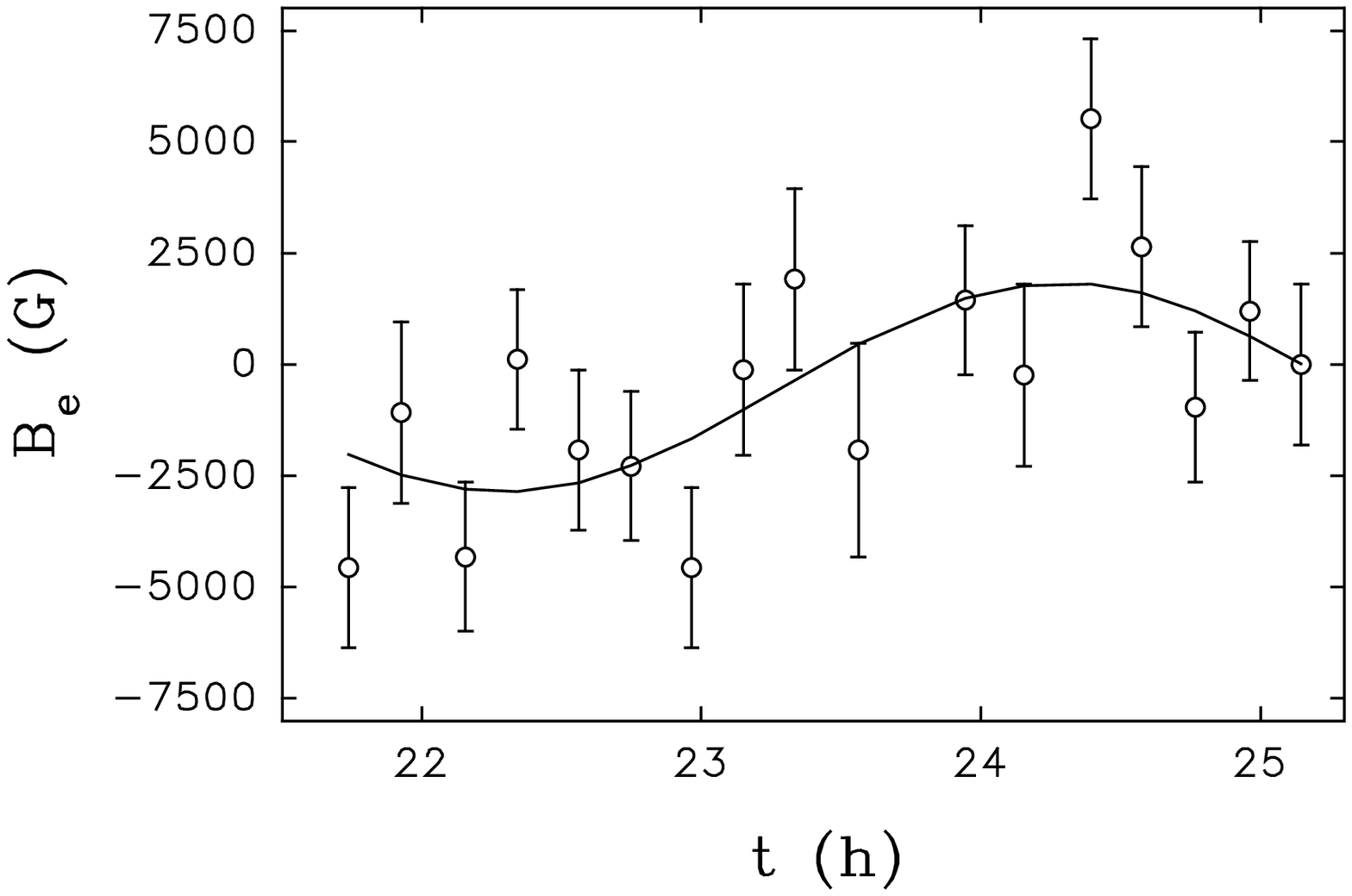,width=7.5cm,height=5cm,angle=0}
\hspace*{1cm}

Fig.~1: The time--resolved Zeeman spectral observations of 40~Eridani~B
with the Main Stellar Spectrograph obtained on December 5, 1995

\end{figure}

The brightest degenerate
star 40~Eri~B was observed in the same mode on September 14, 1995.
The average magnetic field, $B_e = 0.5 \pm 0.37$~kG, during a time of observations of 81~min has been
found. However in another observing run,  on
December 5, 1995,  we found about 4~-hours'
sinusoidal variations of the effective field of 40~Eri~B during  215~min of observations. These observations
are displayed in Fig.~1.
They consist of 54 (3 min) individual measurements averaged in 17 bins.
The best
fit is $B_e = A + B \cos(2\pi t/P + \phi)$ with $A = -510 \pm520$~G,
$B = 2300 \pm 700$~G and a period $P \approx 4$~hours. It was
concluded that for reliable detection of magnetic field and rotational
period in 40~Eri~B such mode of observations must be continued.

The MSS observations of 40~Eri~B might be interpreting not only as the 4~- hours'
periodical variations. The period suspected cannot be considered as
reliable, because the time of observations is comparable with the
period, the scatter of the individual points is high, and one can say only
that a variability of the magnetic field was detected that time. In addition
we  found other periodical signals in those data, one of them was
$\approx 37$~min. There was also a period of about 10~min detected
in the previous
observing run on September 14, 1995. Indeed, we could say that some
probable magnetic field variations were registered that time.

\begin{figure}
\psfig{figure=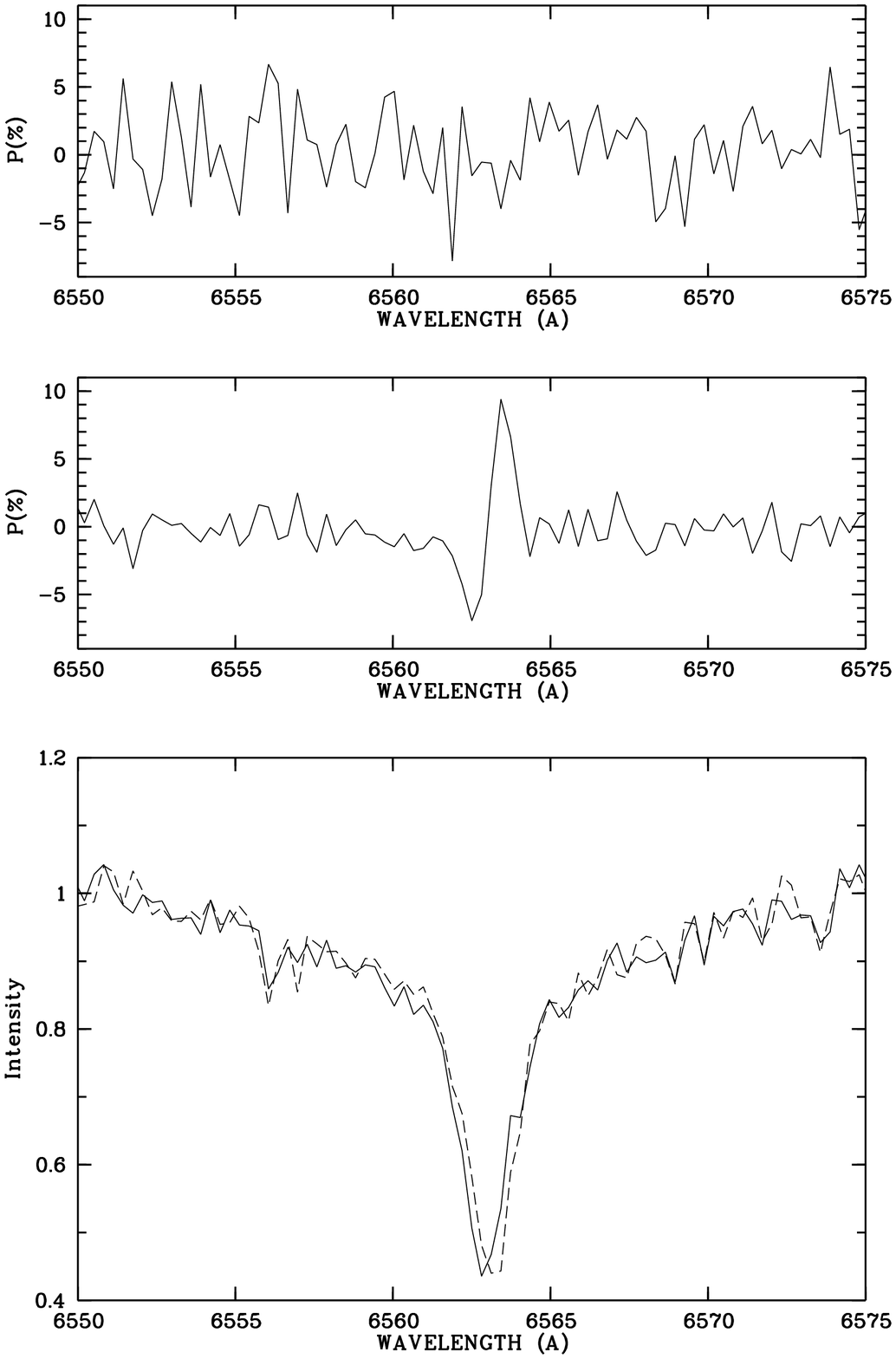,width=9cm,height=11cm,angle=0}
\hspace*{1cm}

Fig.~2: Opposite circular polarization spectra of 40~Eridani~B in the region
of the H$\alpha$ line core and corresponding polarization (bottom and middle)
obtained from the individual spectra (Fig.~1) showing the maximal positive
polarization. Atmospheric water lines around H$\alpha$ are also visible.
At the top we show polarization obtained from the individual spectra
of the zero polarization ``cross--over points''

\end{figure}

In order to demonstrate the magnetic shift of the H$\alpha$ line in
opposite circular polarization spectra, we present here the results of analysis
of two sum double--polarized spectra obtained from the individual
spectra for the December 5, 1995 observing run. The first sum spectrum
consists of
individual spectra showing the maximal positive polarization (the top
points in Fig.~1, situated between the 24th and 25th hours of the time axis),
and the other spectrum consists of the zero polarization spectra
(the ``cross--over points'' in Fig.~1, situated between the 23th and 24th hours).
In Fig.~2 two spectra of different polarization from the first
summed  image in the region of the H$\alpha$ line are shown (bottom). The
magnetic shift in H$\alpha$ is clearly visible. There are some
${\rm H_2O}$ atmospheric absorption lines in the region not showing
a shift. The result of subtraction of these spectra , i.~e. the
polarization (V--parameter) is shown in the middle of Fig.~2. The typical
S--wave polarization in the line profile is detected. At the top of Fig.~2
the polarization obtained from the second (``cross--over'') image is shown.
No detectable polarization is observed there.

\section*{Observations with the SP-124 spectrograph}

In 1997 we changed the observational mode to the medium resolution
spectrograph SP--124. This spectrograph was equiped with a
PM--CCD (1024x1024) providing spectra of a high quality (Neizvestnyi et al., 1998)
and with a new polarimetric analyzer having a better transparency
(Bychkov et al., 2000). The analyzer is equiped with a rotatable
quarter--wave plate.
The spectral resolution in this mode is 5--6~$\AA$ (2.3~$\AA$/pix).
With such a resolution we have lost the possibility of  making measurements of
the narrow non--LTE H$\alpha$ peak, instead we can study two hydrogen
lines H$\alpha$ and H$\beta$ simultaneously, measuring central cores of
these lines. Otherwise the method is the same, the left-- and
right--circular polarized Zeeman spectra are obtained simultaneously
on the detector. The main advantage of this mode is that one can obtain Zeeman
spectra of WDs with high signal--to--noise because of the better CCD cosmetic
and stability (and the lower spectral resolution), besides it provides for recording
 of two hydrogen lines in the spectrum.\\

\begin{table}[h]
\begin{center}
\small
Table 1: Results of magnetic field measurements \\
~\\
\begin{tabular}{|c|c|c|r|r|}
\hline
Target(WD)&   N&    JD&  ${\rm B_e (G)}$& ${\rm \sigma(B_e)}$(G)\\
	   &    &    +2450000&  $ $& $ $ \\
\hline
2032+248&   2&         684.342&   810&      4200 \\
2326+049&   3&         684.379&   150&      2400 \\
0148+467&   2&         684.433&  --390&     4500 \\
1126+185&   2&         950.841&  --3200&    10500 \\
1953--011&   6&        357.479&   28000&    6000 \\
\hline
\end{tabular}
\end{center}
\end{table}

In Table~1 we present new results of magnetic field measurements in 5 DA
degenerates in the moderate resolution mode. N is the number of spectra
obtained. We found no magnetic field in four stars, but a reliable magnetic
field has been detected in WD\,1953---011. In a
massive Zeeman spectroscopy of DA WDs Schmidt \& Smith~(1995) found
a magnetic field, $B_e = 15.1 \pm 6.6$~kG, of WD\,1953--011. This result
suggested WD\,1953--011 to be a magnetic WD candidate. This star was also
suspected as
magnetic by Koester et al.~(1998) in their  high--resolution
spectroscopy observations of the NLTE H$\alpha$ cores in DA WDs. In this program
of  searching for rotation in WDs the narrow H$\alpha$ cores are fitted with
broadened NLTE models, and thus projections of rotation velocities are derived.
In normal DA stars the rotation velocities are extremely small with
typical upper limits for v~sin\,i of about 15~km/s. In WD\,1953--011 a
formal v~sin\,i has been found to be 173~$\pm$~10km/s, however a clear Zeeman
splitting has been detected in two independent spectra. The splitting
corresponds to a surface magnetic field $B_s \approx 93$~kG (Koester
et al.,~1998). In our observations the magnetic field of WD\,1953--011
was firmly detected, $B_e = 28 \pm 6$~kG. In Fig.~3 the spectrum
(the sum of two Zeeman spectra) of WD\,1953--011 is shown
as well as the circular polarization derived from the Zeeman spectra.
The S--wave polarization pattern is clearly seen both in H$\alpha$ and
H$\beta$.

\begin{figure}
\psfig{figure=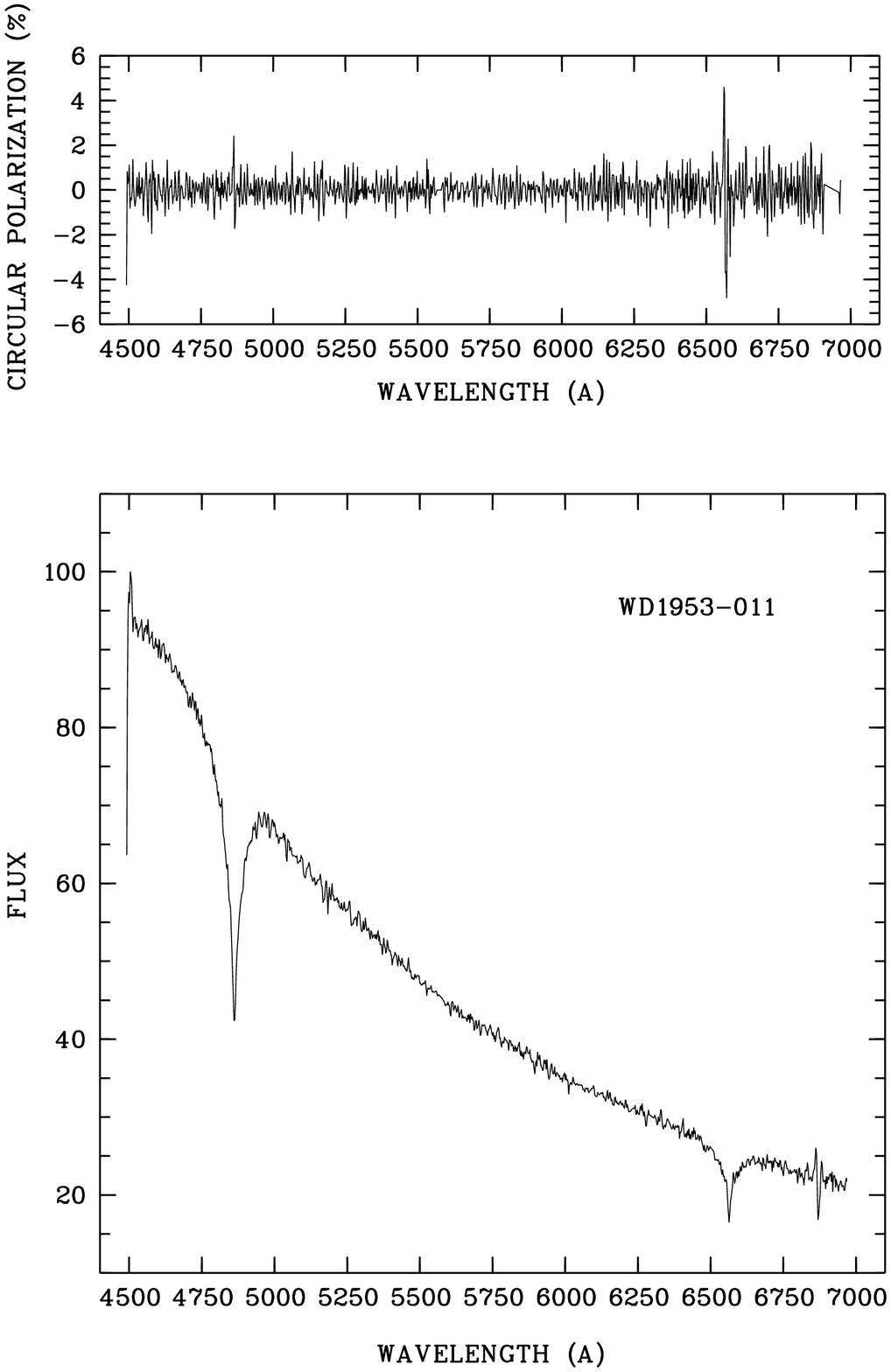,width=8.5cm,height=11cm,angle=0}
\hspace*{1cm}

Fig.~3: A spectrum and curcular polarization of the magnetic degenerate
WD\,1953--011

\end{figure}

The magnetic nature of WD\,1953--011 was thus confirmed  in
direct Zeeman spectroscopy. Koester et al.~(1998) discussed the fact of
considerable systematic difference in magnetic fields estimated in
high--resolution spectroscopy from H$\alpha$ NLTE core splitting and
those found in Zeeman spectroscopy of moderate resolution. This could be
an effect connected here  with  different parts of the line profile
which are measured by these different methods. However it is hardly
probable for a magnetic field to be notably different depending on the
star atmosphere height. The longitudinal magnetic field varies with rotational
phase over wide limits, from 0  to almost the polar field $B_p$,
depending on orientation; at the same time the surface (integral) field
varies with  rotation over much narrower limits. In a model of
a dipolar magnetic field with occasional orientation of the dipole axis
the surface and longitudinal (effective) magnetic fields are related by
a statistical relationship (Angel et al.,~1981) $B_s \approx 3 B_e$.
In this particular star WD\,1953--011 our result, $B_e \approx 28$~kG,
and that by Koester et al.~(1998), $B_s \approx 93$~kG, agree well with
the statistical relation.

The known magnetic degenerate star WD\,0009+501 was observed
on September 1, 1999 in the same mode, the time--resolved Zeeman
spectroscopy. WD\,0009+501
has been discovered as magnetic by Schmidt \& Smith~(1994). In 12
observations for 4 observing nights they found a variable magnetic field
in the range $B_e = +9 \div -100$~kG with $\sigma (B_e) = \pm 8\div15$~kG.
The field obviously varies because of rotation of the star. A power
spectrum analysis has shown a variety of peaks between the 2-nd and 20th hours,
with some preference for shorter periods (Schmidt \& Smith~1994).
There were about 12 minutes
between the spectra in an observational pair, and no variability of magnetic
field was detected inside the pairs, but obvious variations were detected
in adjacent observations separated by about 5 hours. They observed
a nearly full  amplitude of variability (about 90~kG) during an interval of 5 hours.

\begin{figure}
\psfig{figure=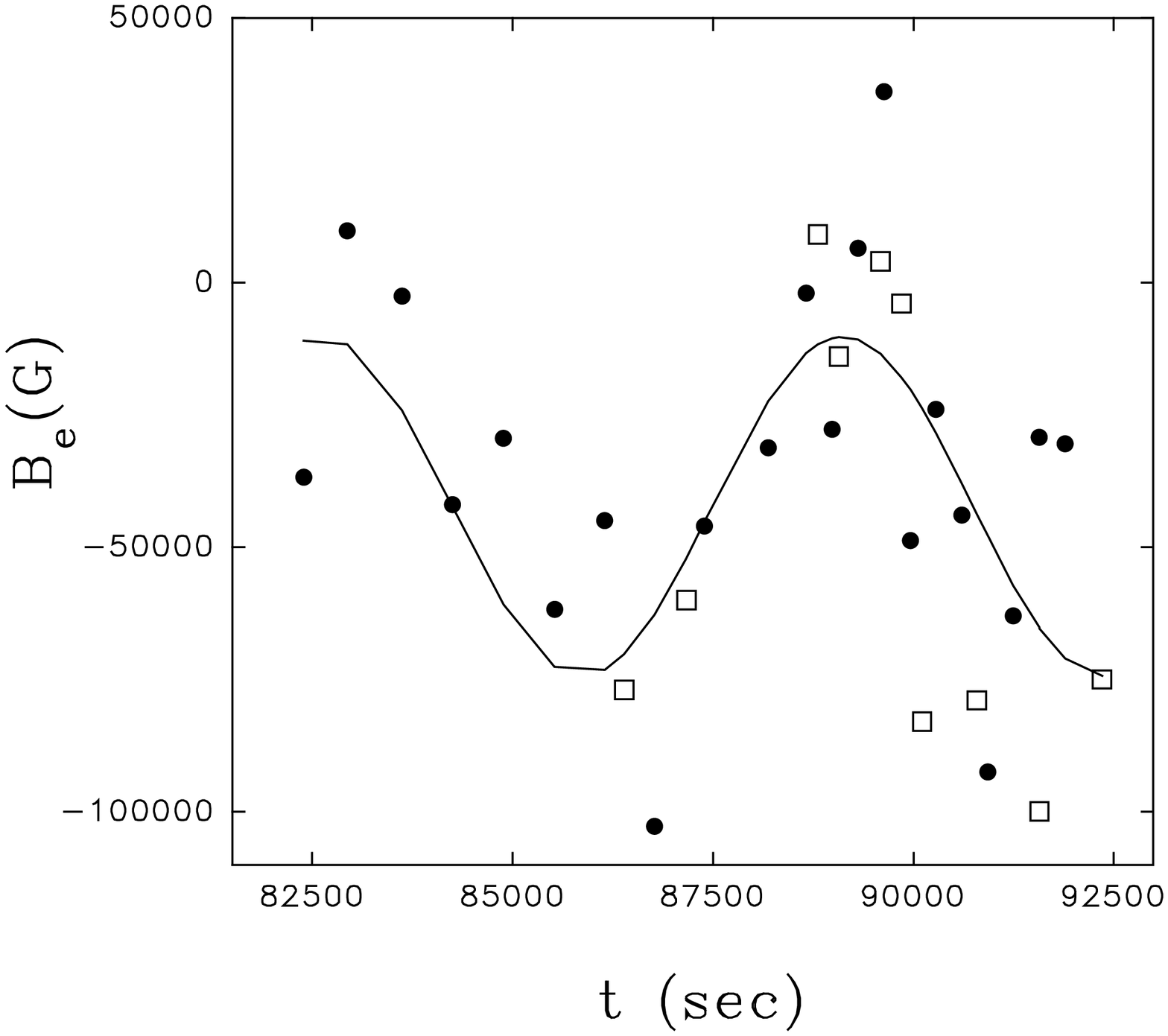,width=8cm,height=6cm,angle=0}
\hspace*{1cm}

Fig.~4:
Time--resolved Zeeman spectroscopy of the magnetic white dwarf
WD\,0009+501. The results by Schmidt \& Smith~(1994) are imposed on
our data, they are indicated by open squares

\end{figure}

Our observing run of WD\,0009+501 consists of 21 continuous Zeeman spectra,
the first 10 spectra were taken
with a 10~-minut of exposure, and the next 11 spectra with 5~-minut of
exposure. Results of measurements of the H$\alpha$ line are shown in Fig.~4.
The total profile of the line in opposite polarization spectra was measured
by a Gauss--analysis. The measurements of H$\beta$ are not shown here,
they are of about the same behaviour, but less accurate. The
observations cover a 2.6~-hours' interval. The variability of magnetic
field is clearly seen in Fig.~4 (filled circles). It was not possible
to derive the period accurately from this observational series alone.
One can conclude from our data only that the rotational period of
WD\,0009+501 is about 2 hours. However with addition of the data by
Schmidt \& Smith~(1994) and analyzing these total data together we
have found the rotational period, it is 1.83 hours. The same
unique period has been identified in their and our power spectra.
In Fig.~4 the 10 close separated in time observations by Schmidt
\& Smith~(1994) are shown as open squares. They have been shifted by
arbitrary phase of the 1.83--hours period. The curve in the figure
is sinusoidal fitting of our data. The average magnetic field is
$<B_e> = -42.3 \pm 5.4$~kG, the semi--amplitude is $32.0 \pm 6.8$~kG.

\begin{figure}
\psfig{figure=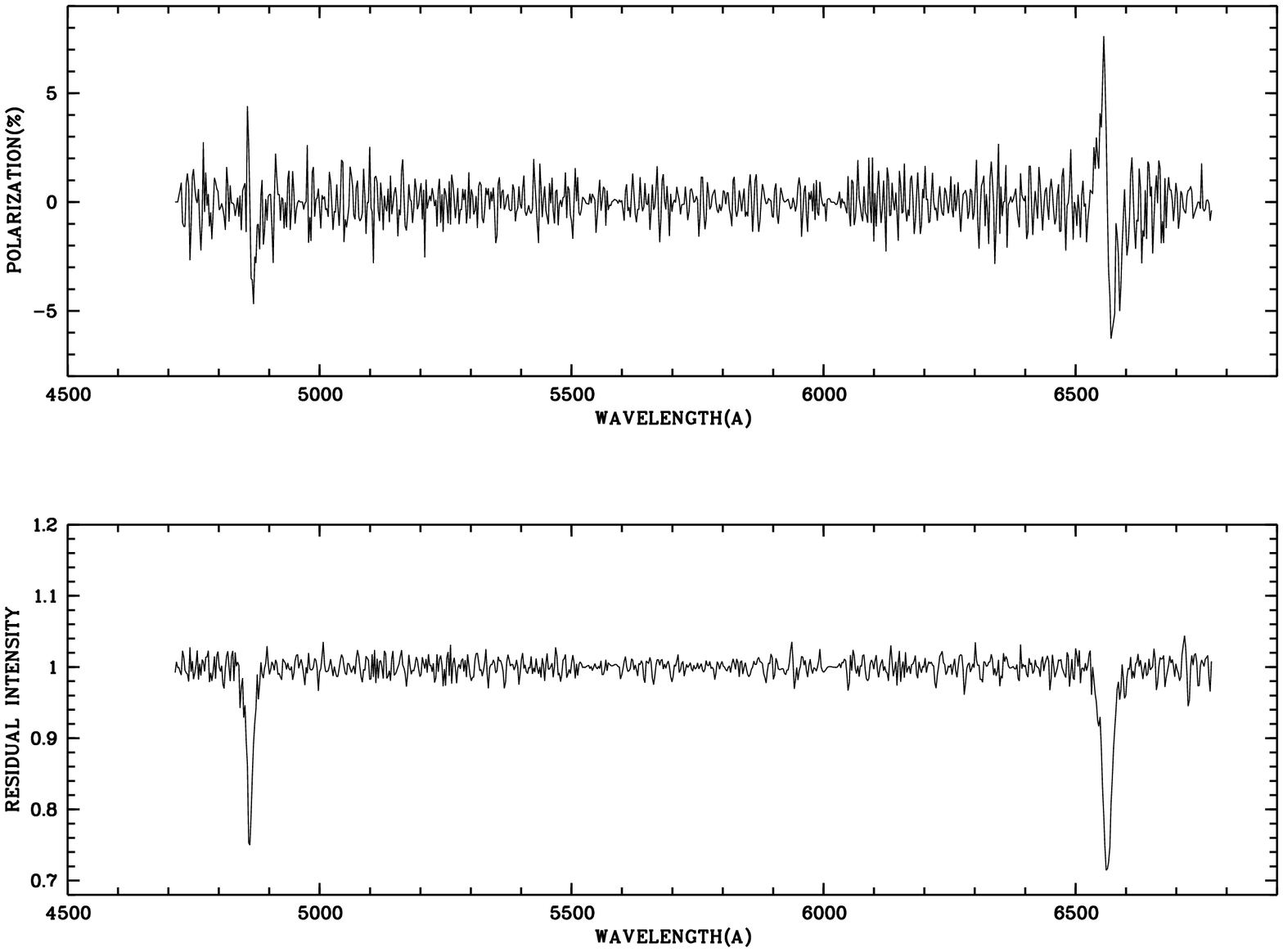,width=9cm,height=7cm,angle=270}
\hspace*{1cm}

Fig.~5:
A spectrum and curcular polarization of the magnetic degenerate
WD\,0009+501

\end{figure}

In Fig.~5 we present both the summed unpolarized spectrum of WD\,0009+501
(bottom) consisting of 8 spectra showing maximal magnetic field
and the corresponding circularly polarized spectrum (top). The
polarization is clearly seen both in H$\alpha$ and H$\beta$.

\subsection*{Magnetic field of 40~Eridani~B}

The brightest DA star 40~Eri~B was observed in the SP-124 mode
 on January 25, 27, and 28 1999 (hereafter~--- the first,
second and  third night).
The aim was to check the magnetic field variability suspected
in our observations in the MSS--mode earlier (see above). A total of 66 Zeeman images  were taken
 (correspondingly 10, 26 and 30 images on these
particular nights) with an exposure time of 5 minutes. On the whole the observations lasted
for 70, 175 and 192 minutes, respectively  on these nights,
but they included the time for rotation of the quarter--wave plate and other
observational operations. We rotated the plate every 5 (sometimes 10)
exposures. This operation is very important for both to take correctly
into account many systematic effects connected with nonideal orientation
of the slit and the polarimetric analyzer and to project the spectra of
opposite polarization onto the same pixels of the detector. Nevertheless,
in the case of magnetic variables like our target 40~Eri~B is, this procedure
alone can not solve the problem of systematic shifts. For this reason
we took spectra of
standard stars in the same mode just before and after the observations of
40~Eri~B. As standards, we used different bright stars which were chosen
depending on a current observational program and situation. These are
the cool stars $\epsilon$~Tau, $\alpha$~Cep or neighbouring stars.
The instrumental line shifts in opposite polarization spectra have been
measured in the standards to control any possible systematic shifts
with  time. In the first star all lines have been measured by the
cross--correlation method in restricted ($\approx 100~\AA$) spectral
regions around H$\alpha$ and H$\beta$. Other standard stars are
of about the same spectral class as the target or have clearly visible
hydrogen lines, so we have measured
these two lines in their spectra. One cannot expect
effective magnetic fields $> 1$~kG in bright stars selected by chance
(Monin et al.,~2000), and all these stars could serve as standards
to control our ``magnetic zero'' variability with an accuracy
of $\approx 1$~kG or better. Instrumental shifts of about 1~--3~kG were really detected
in these observations for 2~-- 3
hours of observing, and the
data were corrected for these shifts by a linear interpolation.

%
%

Two hydrogen lines H$\alpha$ and H$\beta$ have been analysed in 40~Eri~B.
All the spectra have been normalized with the same
window, 180~$\AA$. Relative shifts in the circular polarized spectra
have been measured by the Gauss--analysis method. Only central
line cores of about 30~$\AA$ in width were thus measured in the
normalized spectra.
Both the normalization procedure and the spectral filter parameters were
optimized and they were used in a strictly  the same way in all the spectra.
Not all the spectra we obtained  were finally used in magnetic field
measurements. The target is low enough at the 6--m telescope latitude
(the zenith distance is $51^{\circ}$ at the culmination). We observed
40~Eri~B around the culmination and tried to obtain the observations
as long as possible, nevertheless, depending on weather conditions
not all the spectra were obtained with a desired quality
(${\rm S/N >80 \div 90}$ in an individual spectrum of one polarization).
We selected the best spectra {\it before} the measurements of magnetic
field, and only the ones where the hydrogen line cores were not
distorted by cosmic ray particles. Particle traces in the steep line
cores, even being corrected,  may result in distortion of the line shifts.

Results of magnetic field measuremets in H$\alpha$ line from
individual Zeeman spectra of
40~Eri~B are shown in Fig.~6 for the first (top) and the third (bottom)
nights of the January 1999 observing run. The crosses indicate the magnetic field
of the target, the stars indicate that of the standards. On the first night
the stars $\alpha$~Cep (before) and $\epsilon$~Tau (after) were observed
as standards. On the third night there were neighbouring stars, a star
situated about $19^{\prime}$ to north (before) and 40~Eri~A (after). The first
star  was unfortunately not bright enough on the last night to provide
a desired accuracy of the zero--field point ($< 1$~kG). This may result in
some uncertainty as a total shift (a linear trend) in calibration of
the first points in this series of observations. One can see the variable
magnetic field in 40~Eri~B detected in H$\alpha$.

\vspace{1cm}
\vbox{
\psfig{figure=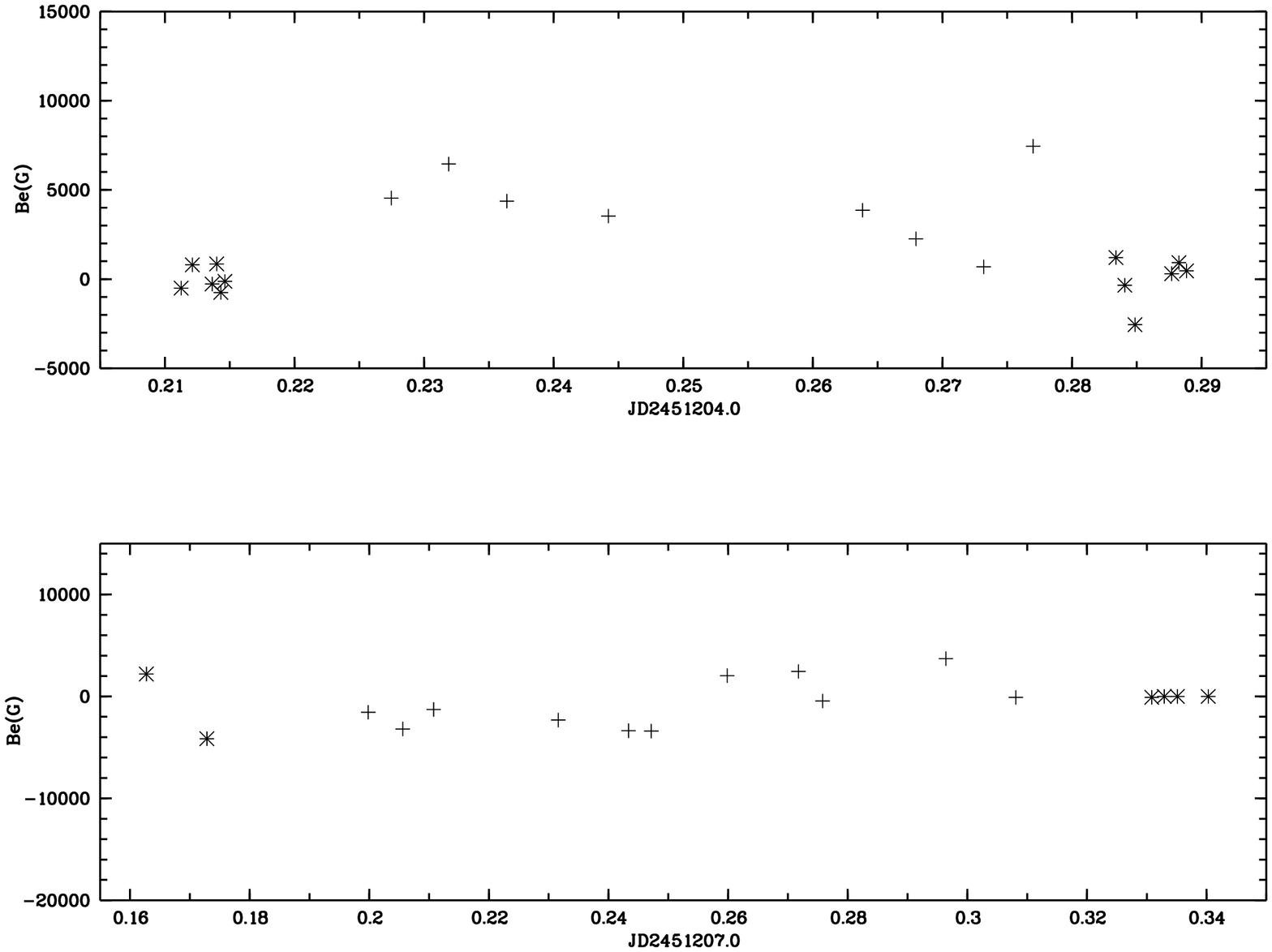,width=9.3cm,height=8cm,angle=270}

\hspace{-0.65cm}
Fig.~6:
The H$\alpha$ line magnetic field variability of 40~Eridani~B
obtained on the first and third nights of observation
}

\vspace{1cm}
Power spectrum analysis of the data from all three nights together,
both Fourier and a least--squars search for sinusoidal signals
display several peaks from about 2 hours to 5~-- 6 hours. These peaks
are not independent, they are on a common origin consisting in
variable, periodical and probably, not sinusoidal signal.
Magnetic phase curves in H$\alpha$ and H$\beta$ of two best periods,
${\rm 2^h\,25^m}$ and ${\rm 5^h\,17^m}$, are displayed in Fig.~7 and 8
respectively. Both the H$\alpha$ and the H$\beta$ lines confirm the periodical
variations, they are about the same both in phase and in amplitude,
though the scatter in the H$\beta$ data is greater. The individual data are
presented by open circles, the mean values averaged in the phase bins
$0.1 \Delta \phi$ with their r.m.s. errors are shown by filled circles.
We present  in these figures also the sinusoidal fitting of the data
as $B_e = A + B \sin(2\pi t/P + \phi_0)$. The phase $\phi_0$ was
fixed the same in  both lines. The parameters of the curves have been
found as follows.\\

\vbox{t}{
\centerline{\hbox{
\hspace{-0.3cm}
\psfig{figure=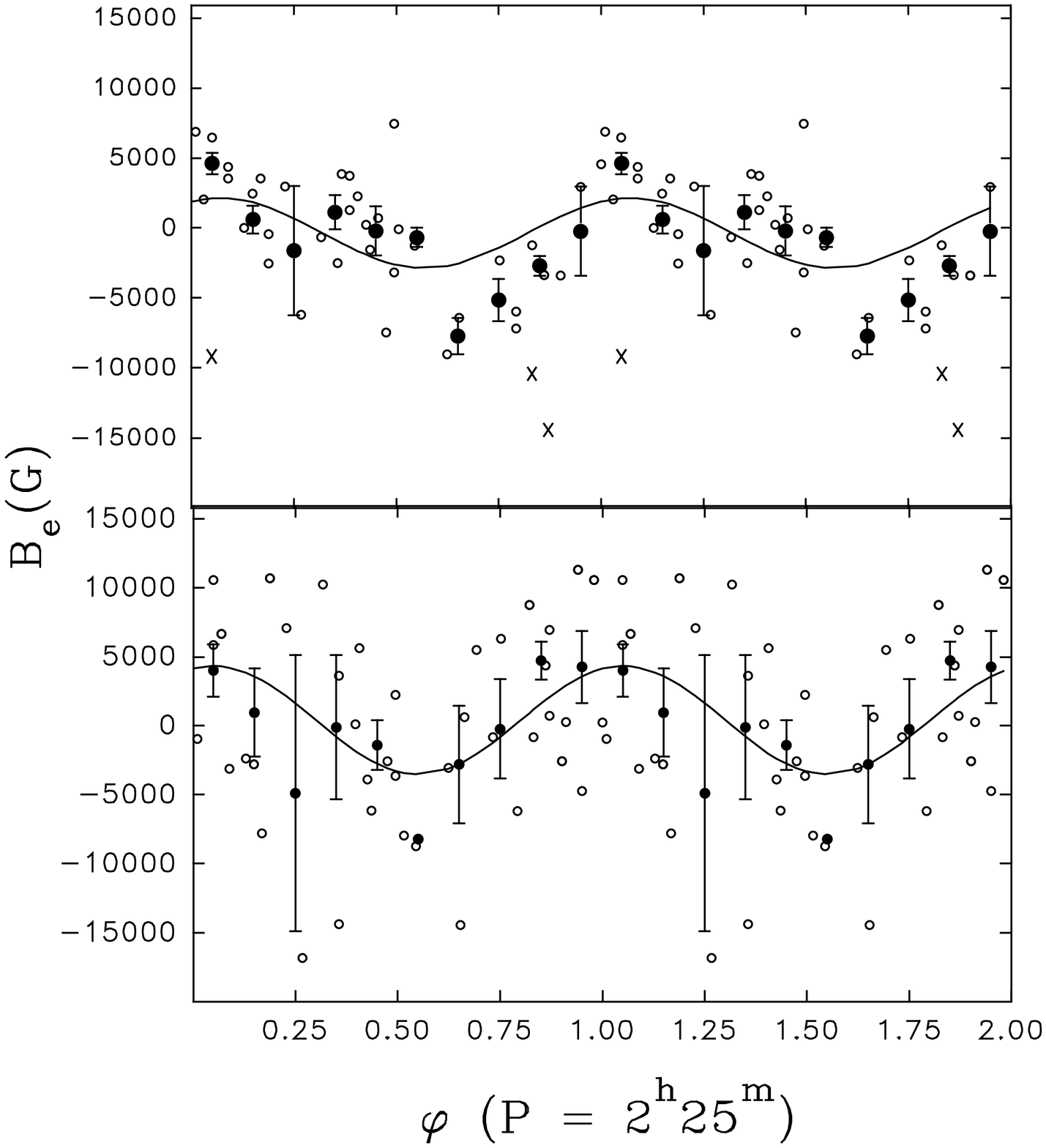,width=8.5cm,height=8cm,angle=0}
}}

\hspace*{-0.65cm}
Fig.~7:
The first period magnetic field phase curve for H$\alpha$ (top)
and H$\beta$ (bottom)

\vspace{1cm}

\centerline{\hbox{
\hspace{-0.3cm}
\psfig{figure=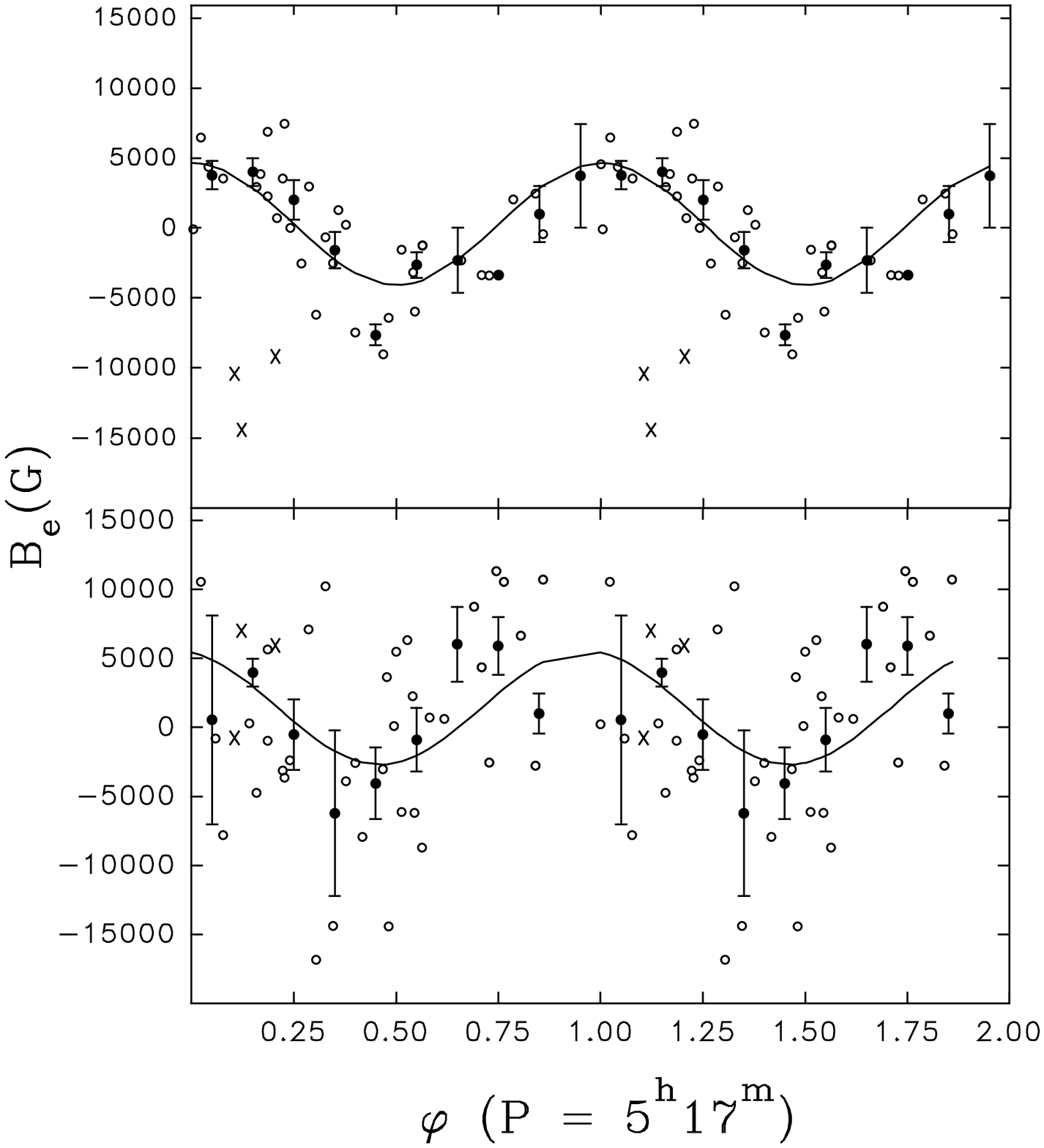,width=8.5cm,height=8cm,angle=0}
}}

\hspace*{-0.65cm}
Fig.~8:
The same as in Fig.~7 for the second suspected\\ period 
}

\vspace{1cm}
\noindent
$P={\rm 2^h\,25^m}$  \\
H$\alpha$: \hspace*{0.1cm} $A=-300\pm550$~G,\,\,$B=2500\pm900$~G, \\
H$\beta$: \hspace*{0.1cm} $A=400\pm1000$~G,\,\, $B=4000\pm1400$~G,\\
$P={\rm 5^h\,17^m}$\\
H$\alpha$: \hspace*{0.1cm} $A=300\pm550$~G,\,\,$B=4400\pm700$~G,\\
H$\beta$:  \hspace*{0.1cm} $A=400\pm900$~G,\,\,$B=4000\pm1400$~G.\\

\noindent
A number of possible systematic shifts (for instance, the
atmospheric dispersion) may influence the results changing
the positiones of these two lines  in antiphase, because the lines are situated
at different edges of the spectral range covered in our spectroscopy. The
correlated variability means that systematic effects do not distort
the results notably. However 3 points do not agree with the H$\alpha$ magnetic
curves, but agree well with those in H$\beta$. They are marked by the crosses
in Figs.~7,~8. They belong to the second night. We can not exclude here
the influence of systematic effects.

The setting the equalizing  of the zero phases to be equal in H$\alpha$ and H$\beta$
resulted in the sinusoidal fits parameters are not best. For instance,
the H$\alpha$ magnetic field amplitude in the first period has been
found to be too small in the formal fit. The zero phases are really slightly
different in H$\alpha$ and H$\beta$, and this is due to the
individual data scatter. We conclude that the  H$\alpha$ and H$\beta$
data both show the same periodical variations of the effective magnetic field.
The semi--amplitude of the variations, $B_{max} \approx 4000 \div 5000$~G, and
average field is about zero $\pm 500$~G. The direct average magnetic
field derived from the H$\alpha$ data only (without the 3 measurements
indicated by crosses in the figures) is $<B_e> = -200 \pm 500$~G.
The ratio $B_{max} / <B_e>$ is found to be very high. From  the data in Fig.~7, 8One can make a conclusion
 on the star
and its magnetic field orientation. If the magnetic field of
40~Eridani~B is a central dipole, then the rotational
axis inclination to the line of sight is high, $i \sim 90^{\circ}$, and
the magnetic axis inclination to the rotational axis is about the same,
$\beta \sim 90^{\circ}$.

We may conclude that these new data confirm the previous result
obtained in the 1995 observations with MSS, where we found a variable
magnetic field in 40~Eri~B with a semi--amplitude of $2300 \pm 700$~G
changing  on a time--scale of about 4~hours. However we cannot
determine firmly a real sole period of the variability. We could select two
possible periods in the variability. The first one (${\rm 2^h\,25^m}$)
was determined with a much better accuracy than the second one
(${\rm 5^h\,17^m}$). The
latter period is longer than the  longest observational series on
the second and third nights (none of the observations covers the whole 5 hours'
period, they appear as fragments
of the magnetic phase curve in Fig.~8). Nevertheless, the 5
hours' phase curve demonstrates that if the rotational period of 40~Eri~B
is $\gea 5$~hours, the magnetic curve must be non--sinusoidal
(a non--dipolar magnetic field).

\begin{figure}
\psfig{figure=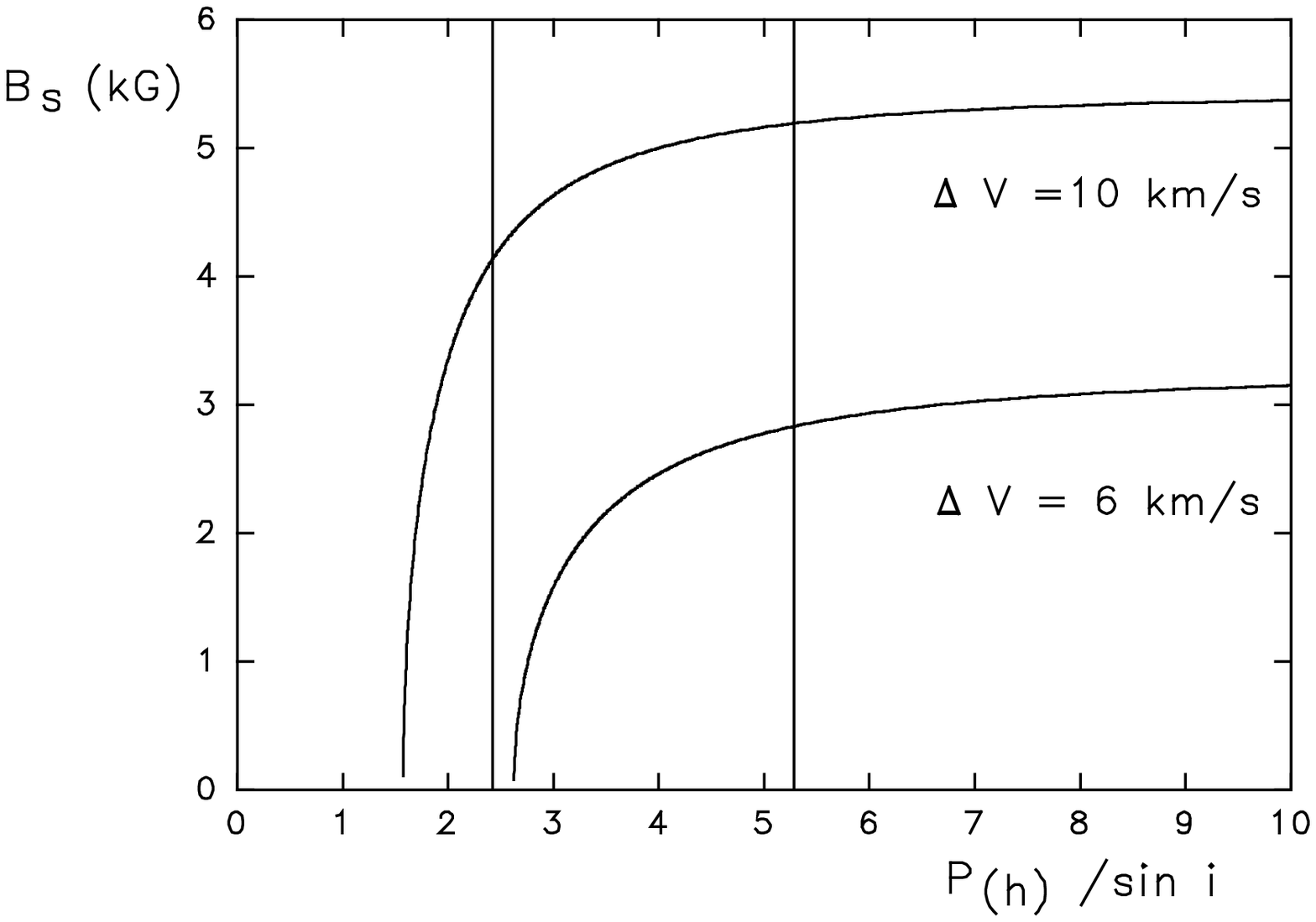,width=8cm,height=6.5cm,angle=0}
\hspace*{1cm}

Fig.~9:
The relation between surface magnetic field and
rotational period of 40~Eridani~B. The curves correspond to the
upper limits on the total width of the non--LTE H$\alpha$ core, vertical
lines indicate the two rotational periods derived in this paper. The
permitted region is located below upper curve and to the right of the
left vertical line

\end{figure}

The rotation and the magnetic field could be tested together through
their impact on the line broadening. The two effects broaden the
central core of the H$\alpha$ line independently. The surface magnetic field
splitting is ${\rm \Delta V_{mf} = 1.84\, B_s}$\,km/s\,kG. The rotational
broadening is ${\rm \Delta V_{rot} = V_{rot} \sin\,i = 4 \pi R/(P/\sin\,i)}$,
where R is the radius and P is the period. The total broadening
${\rm \Delta V = \sqrt{\Delta V_{mf}^2 + \Delta V_{rot}^2}}$ is the
observed quantity. In high--resolution spectroscopy the central H$\alpha$
core profile being fitted to broadened NLTE models
yields  an estimate of the velocity ${\rm \Delta V}$. Heber et al. (1997)
presented such a study of 40~Eri~B; they found the broadening of less than
${\rm \Delta V}< 8$~km/s to be the upper limit at a $3 \sigma$ level. We show in Fig.~9 the
relation between ${\rm B_s}$ and ${\rm P/\sin\,i}$ for this star
, where the radius ${\rm R = 8.98 10^8}$~cm (Reid, 1996) is
 accepted. Two curves in the figure indicate the total broadening
${\rm \Delta V = 10}$ and 6~km/s.

The permitted region for the magnetic field and rotation of 40~Eridani~B
is one below the curves.
We present these two curves keeping in mind that the magnetic
broadening can be variable with rotational phase. It varies
depending on the dipole parameters and on the unknown phase of
the suspected (2~--5 hours') rotational period, when the spectra for the
$v \sin i$ analysis were taken (Reid, 1996; Heber et al., 1997).
The surface magnetic field strength may change by a factor of $1 \div 2$ in
the course of rotation. This estimate follows from the study of magnetic
Ap stars by Mathys et al. (1997). They collected data on 42 well--studied
magnetic Ap stars, their magnetic curves and the mean magnetic field modulus.
The ratio $q = <B>_{max}/<B>_{min}$ of the observed maximum and minimum
values of the mean surface magnetic field modulus varies from star to star
over the limits indicated above. About 60\,$\%$ of the stars have $q = 1.0 - 1.1$,
and the other stars have this ratio $q = 1.1 - 1.9$. Taking into account
the magnetic broadening as a probable variable contributor to the total line
broadening and ${\rm \Delta V}< 8$~km/s as the observed $3 \sigma$
upper limit (Heber et al., 1997), we can adopt ${\rm \Delta V}< 10$~km/s
as a very probable upper limit of the H$\alpha$ core width in this star.
Two vertical lines in Fig.~9
correspond to the rotational periods ${\rm 2^h\,25^m}$ and
${\rm 5^h\,17^m}$ (at $\sin i = 1$), which we discussed above. The permitted
region for 40~Eridani~B is that both below the upper curve and to the right
of the ${\rm 2^h\,25^m}$ vertical line. We can conclude that the
high--accuracy Zeeman observations must be continued to clarify the rotation
of the star and its magnetic field curve. Fig.~9 demonstrates the
possibilities. In the near future we will be able to know both the rotational period and
the magnetic field structure and orientation in 40~Eridani~B.

\begin{figure}
\psfig{figure=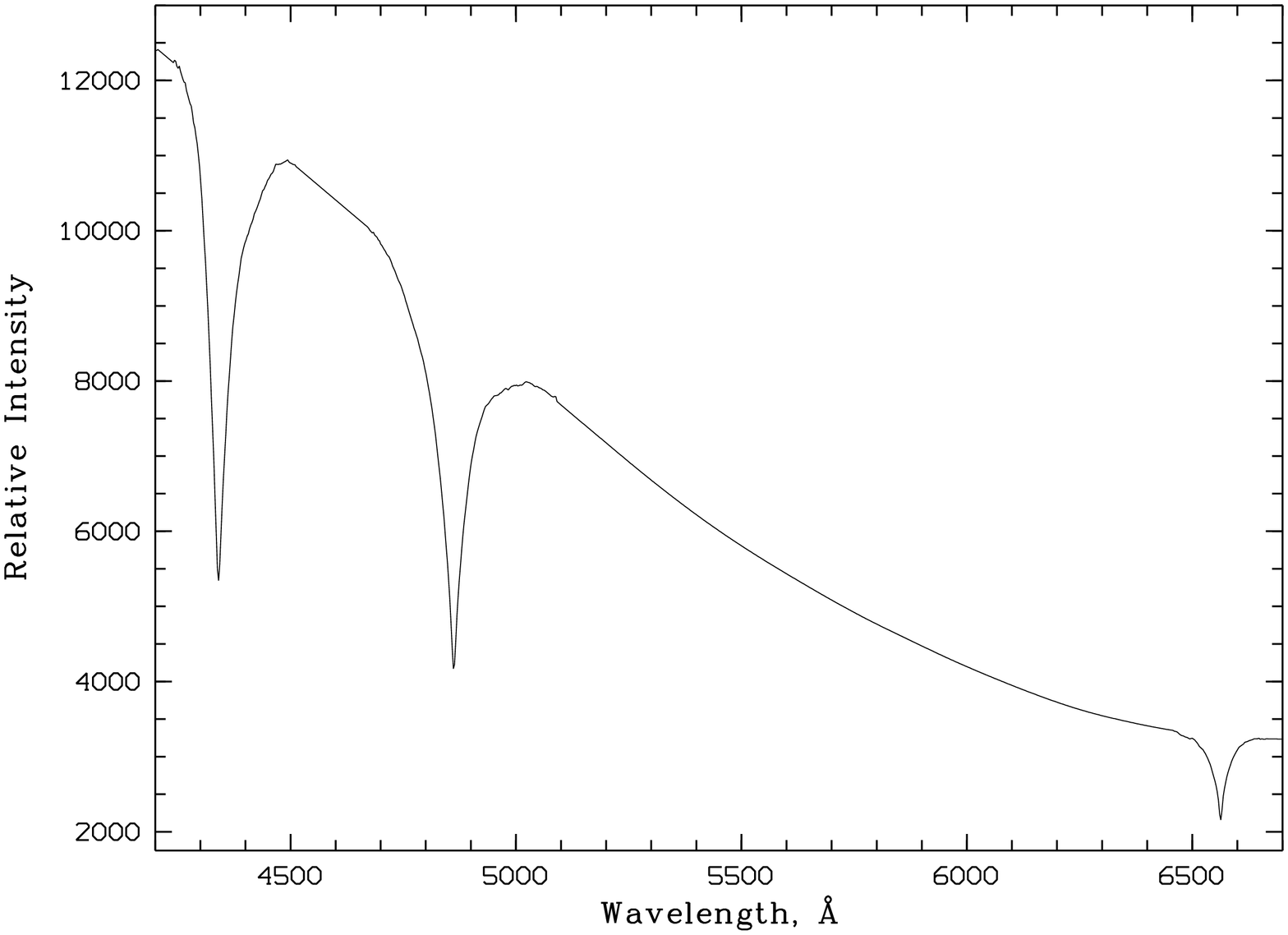,width=8cm,height=6.5cm,angle=270}
\hspace*{1cm}

Fig.~10:
The total spectrum of 40~Eridani~B obtained from 3 nights
\end{figure}

\subsection*{Spectrum of 40~Eridani~B with ultrahigh signal--to--noise}

The presence of heavy elements in the
atmospheres of very hot DA (H--rich) white dwarfs is well established. These are a group of
stars with effective temperatures in excess of 55000~K (Feige~24,
G\,191--B2B). Being originally discovered with IUE (Bruhweiler \&
Kondo, 1981), the heavy elements are extensively studied in UV and far
UV spectra of the hottest DAs, which show the presence of absorption
lines of C, N, O, Si, S, P, Fe and Ni (Sion et al., 1992; Barstow et al,
1993; Vennes et al., 1996). The heavy elements are separated to the
photosphere by the radiative levitation.
There is evidence for the stratification of Fe in the atmosphere of
G\,191--B2B (Barstow et. al., 1999), Fe is stratified with increasing
abundance at greater depth. Stratification of elements
is obtained self--consistently in atmospheric model atmospheres with account for
gravitational settling and radiative levitation (Dreizler \& Wolf, 1999).

Cool DAZ degenerate stars are believed  to accrete
interstellar gas, which enriches their photospheres. A few DA stars
display Ca, Mg, Fe lines both in the UV and in visible regions. They are
G\,74--7 (7300~K) (Lacombe et al., 1983; Billeres et al., 1997),
G\,29--38 (11000~K) (Koester et al., 1997), G\,238--44 (20000~K)
(Holberg et al., 1997). Some lines (Mg\,II~$\lambda$\,4481)
originating from the excited lower level demonstrated that they cannot arise
in the interstellar gas.
Recently Zuckerman \& Reid~(1998) have observed 38 cool DA stars
with HIRES on the Keck~I telescope. They have searched for the Ca\,II~K line
which could indicate a gas accretion. They have found
that the CaII~K line was registered in  spectra of about 20~$\%$ of the stars.
In this situation the border between DA and DAZ stars looks rather
illusory.  It is very important to get more information on the signs
of the elements in the atmospheres of white dwarfs both from UV spectra and
in visible ones using very high signal--to--noise spectroscopy.
In the controversy between line--free DC and DA/DB stars such a
spectroscopy has drastically reduced the number of DC stars (Greenstein, 1986).

In the medium--temperature atmospheres $\lea 20000$~K, as it is in
40~Eridani~B, the radiative pressure in not so strong as to separate
heavy elements and to enrich the photosphere. As it
follows from the simulations by Chayers et al. (1995) of expected
equilibrium abundances of heavy elements levitating up to the photosphere,
in DA stars with a temperatures $\lea 20000$~K
one may expect Al/H and Si/H~$\lea 10^{-9}$, and other elements
are less levitating at such low temperatures. The theory predicts
all heavy elements in the photospheres of DA white dwarfs have to settle down
by gravitational sedimentation, which is a very rapid and effective process.
For instance, the settling time for metals at a temperature of 15000~K
is about 3--4 days, and it is only 0.6~day for He (Paquette et al., 1986).

Another possibility of supplying the photospheres of DAs with heavy elements is
a gas accretion (Alcock \& Illarionov, 1980a; 1980b) from interstellar
medium (ISM) or stellar winds from a companion in binaries. Accreting
mass rate depends strongly on differential velocity $V$ between a star
and the local ISM, $\dot{M_a} \propto V^{-3}$. The distribution of the
differencial velocities between DA stars and the ISM is asymmetrical
(Holberg et al., 1999), the distribution is nearly uniform between
+\,80 and --\,20~km/s. This assumes that the lines measured are related
to the stars or the local ISM disturbed by stellar winds. Basing on the central
part of the distribution, it can be suggested that the relative velocities
may be $\gea \pm \! 20$~km/s. The interstellar gas accretion rate is
$\dot{M_a} = 4 \pi m_H \hat n (G M)^2 V^{-3} \approx 3.2\, 10^{-16}\,
\hat n_1 m_{0.6}^2 v_{20}^{-3}\, M_{\sun}$/y, where the star's mass is
$m_{0.6} = M/0.6\, M_{\sun}$, the relative velocity is $v_{20} = V/20$~km/s,
the number density of the ISM (out of dependence on its ionization state) is
$\hat n_1 = \hat n/1{\rm cm^{-3}}$.

With regard to the interaction of stars with the ISM any star can be
in one of the two states: i) the mass loss and ii) the ISM gas accretion.
The mass loss of hot stars is much more effective than the radiative
pressure in prevention of accretion. However in the case ii) where the star
does accrete a gas, under some conditions
the accretion may be eventually stopped at the
border of magnetosphere and the gas can be expelled by the rotating magnetic
field~--- the propeller (Shvartsman, 1970; 1971; Davidson \& Ostriker, 1973;
Illarionov \& Sunyaev, 1975). Even if a magnetosphere is in the
propeller regime, some portions of the gas may penetrate inside the
magnetosphere and reach the stellar surface due to various plasma
instabilities. A study of tracks of the elements in white dwarfs may help to
understand their status of  interaction with the ISM.

\begin{figure*}
\centerline{\psfig{figure=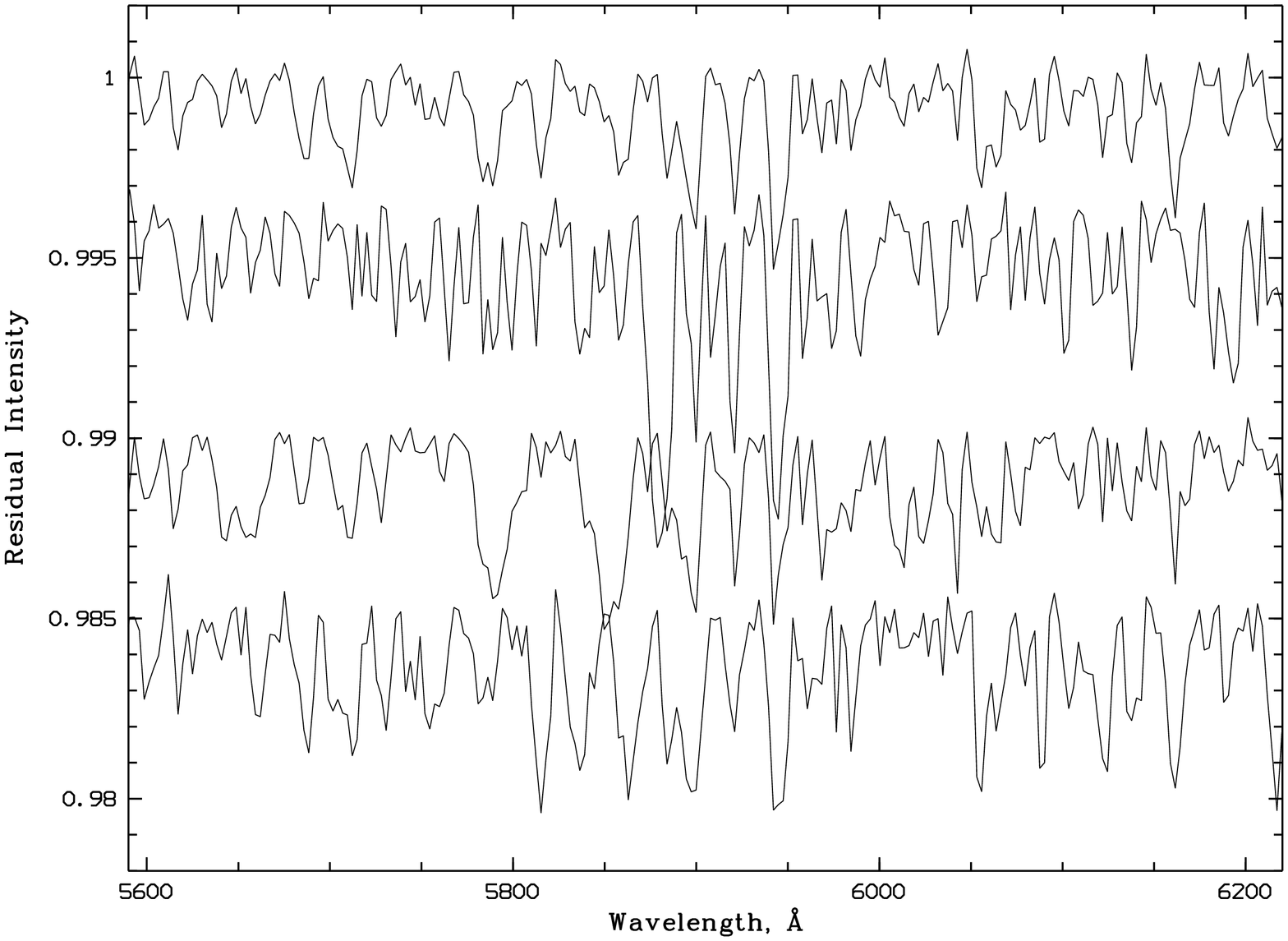,width=19cm,height=8cm,angle=270}}

Fig.~11:
The normalized sum spectra of 40~Eridani~B obtained from
bottom to top~---
in the 3-rd, 2-nd, 1-st nights and the total sum spectrum. The spectra
are consequently shifted on 0.005.
\end{figure*}

\begin{figure*}
\centerline{\psfig{figure=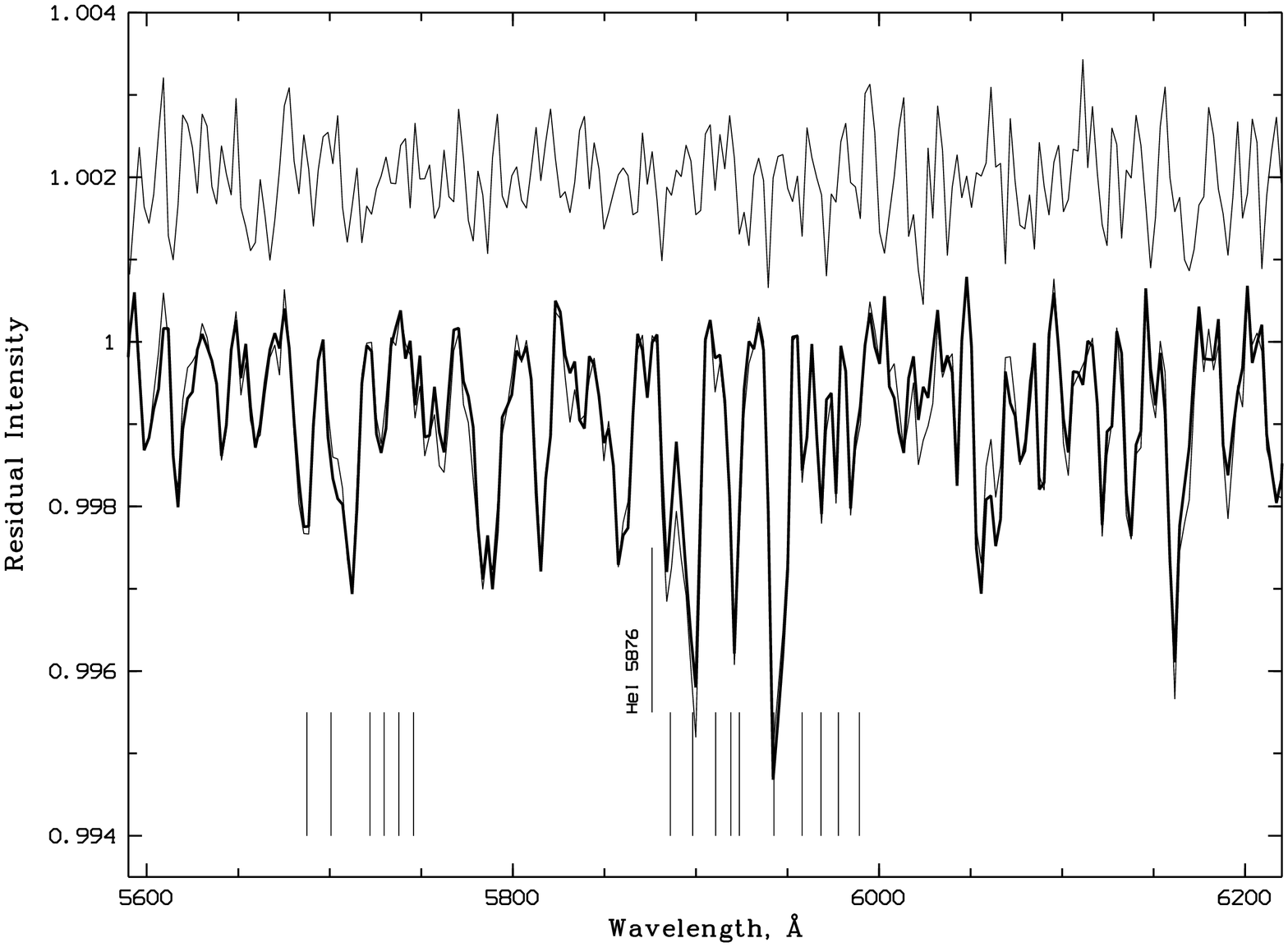,width=19cm,height=8cm,angle=270}}

Fig.~12:
The normalized sum total
spectrum of 40~Eridani~B
(bold line), that without background substruction (thin line) and the
total background spectrum (shifted up). The sky background spectrum was
divided into 20 for better visual inspection
\end{figure*}

The photosphere temperature of 40~Eridani~B is $T_e = 16500$~K.
The radiative levitation cannot supply the photosphere with heavy
elements, but accretion can. If the magnetic field discussed above
is dipolar, it can prevent the accretion depending on the relative velocity
$V$ between the star and the ISM.
Greenstein (1980) reported a discovery with IUE an of absorption
line near $\lambda 1391$ with an equivalent width of 3~\AA, which could be
Si\,IV or, wich is  most probable, the (0,\,5) transition of the H$_2$
Lyman band. The last interpretation suggests  presence of  cool
gas in the upper atmosphere of the star with a column density
${\rm N(H_2) \sim 10^{15}\,cm^{-2}}$ (Greenstein, 1980). In spite of
this the line was not confirmed in the later IUE study
(Bruhweiler \& Kondo, 1983),
the same feature has been reported to exist in other white dwarfs (Wegner, 1982).
No stellar photospheric features has been reported by Holberg et al. (1998)
in the IUE total co--added spectrum of 40~Eridani~B. Only weak
interstellar features due to Si\,II\,$\lambda\,1260$,
C\,II\,$\lambda\,1334$, and O\,I\,$\lambda\,1302$ have been detected there.

We have a good chance to check the theory and to search
for the tracks of elenents in 40~Eridani~B in the visible range.
The total exposure time over the 3 nights of the Zeeman observations
was 5 hours. We shifted the grating by 30~-- 50~\AA \, on each night
in order to avoid the nonuniform  pixel--to--pixel sensitivity of the CCD.
The superflat, superbias and superdark images were prepared and
standard procedures of spectrum reduction were applied. We  obtained
a summed unpolarized spectrum of the star (Fig.~10).
In the range 5500~--~6000~\AA \,
about $4 \cdot 10^6$~counts  obtained in the total spectrum, and
 about $1.5 \cdot 10^6 \div 2 \cdot 10^6$~counts in the blue $< 4500$~\AA \, and
red $> 6500$~\AA \, ranges. The spectrum in the region $5600 - 6200$~\AA \,
is shown in Fig.~11. There are 4 spectra~--- for the 3-rd, 2-nd, 1-st nights
and the total spectrum in the figure from bottom to upwards accordingly.
The spectra were smoothed and normalized with a window of about 50~\AA,
so any information on broader spectral details was lost in the
spectra. THe spectral resolution in the summed spectra is comparatively low,
6~\AA. Unfortunately the Ca\,II K line was beyond the spectral range
registered (the main goal of the spectroscopy was to study
the magnetic field, and hydrogen H$\alpha$ line was the most important).

One can see
a number of weak absorption lines in the spectrum, and the strongest
of them  repeatedly appeared in the spectrum of each night. In Fig.~12
we present the final background--subtracted spectrum (bold line) and
that with no background subtraction
(thin line) in the same spectral range, as well as the corresponding
total background spectrum (top) which has been divided into 20
and shifted up for the best visual inspection. A special study
of the background subtraction was carried out. We found that
any possible errors in the background normalization and subtraction
could not change the resulting absorption lines found in the
40~Eridani~B. The Moon phases were 0.7~-- 0.85 during these observations,
and the background contribution in the spectrum was $\approx 1~\%$.

In Fig.~12 are also shown positions of telluric lines
(Curcio et al., 1964), they are either positions of the strongest line
in the group of lines or positions of groups of lines convolved
with our spectral resolution. In spite of the rather low resolution we
certainly observe these lines in the spectrum. The expected
position of the strongest He\,I line 5876~\AA \, is also shown.
This line intensity is less than 0.001. A preliminary inspection of
this range and other ranges of the spectrum has shown that we observe
numerous weak lines of Ca\,I, Na\,I, Fe\,I, Fe\,II, Mg\,I, Mg\,II, Si\,II at a level of intensities of $0.1 \div 0.5~\%$.
The ultraweak absorption spectrum of the star corresponds to a
temperature of 5000~-- 8000~ë, so they are not photospheric features.
These lines could be interstellar by origin, but not all of them,
because many lines have excited the low level, and Na\,I ${\rm D_1, D_2}$ lines,
for example, are not the strongest among other lines observed.
An analysis of these absorption features will be made elsewhere.
The signal--to--noise in the total spectrum is in the range
1000~-- 2000, but the correct value can be found only after detailed
identification of the spectrum.

We conclude preliminarily that a rich absorption spectrum of heavy
traced elements is present in 40~Eridani~B. Assuming that the photospheric
He\,I line are broadened to FWHM~$\approx 20$~\AA\, we find that
its equivalent width is $W_{\lambda} < 10$~m\AA. In a weak absorption
line approximation, $W_{\lambda} = \pi e^2 {\lambda}^2 N f / m c^2$,
the upper limit of the He\,I line intensity gives
the upper limit of  the column number density of this element $
N(HeI)< 10^{11}~{\rm cm^{-2}}$ or He abundance
$ N(He)/N(H) < 10^{-7}$. This confirms the theory that He
has indeed  diffused under the photosphere in the hydrogen--rich degenerate.

The traced absorption spectrum does not
agree with the photospheric temperature; this suggests that
we have observed either i) the uppermost atmosphere gas in this star, or
ii) circumstellar gas, or iii) interstellar gas.
We believe that the traced heavy elements
are supplied through the gas accretion process, this gas may come both from
the interstellar medium and from the comparatively close companion 40~Eri~C.
The local ISM consists of a few fluxes (Bochkarev, 1990) which move
with some dispersion in about similar directions. In the heliocentric
frame the average velocity of the fluxes is $20 \div 30$~km/s, and the
common direction is $\alpha \sim 90^{\circ},
~\delta \sim 0^{\circ} \div 10^{\circ}$.
This direction is not far from that to 40~Eridani~B. Comparing with
the real radial velocity of the star (Reid, 1996), $V_r = -44$~km/s, we find
that the relative velocity is  high enough. Direct measurements of the ISM
line of sight velocity have been obtained by Holberg et al. (1998),
+\,7~km/s; this gives the differential radial velocity between the
star and the ISM as 51~km/s. In a tangential direction the proper motion
corresponds to a velocity of 94~km/s. We can find that the total relative
velocity of 40~Eridani~B and its local ISM is $V = 100-120$~km/s.
Accepting a density of the ISM close to the Sun $\hat n_{0.1}  =
\hat n/0.1{\rm cm^{-3}}$, find the mass accretion rate $\dot{M_a} \approx
2.2\, \cdot 10^{-19}\, \hat n_{0.1} m_{0.6}^2 v_{100}^{-3}\, M_{\sun}$/y.

The companion of 40~Erinadi~C is an M4e dwarf separated from it by the $a = 40$~a.e.
orbit. The companion's wind is captured by the white dwarf, and the
capture radius is $R_c \approx 2 G M / (V_{orb}^2 + V_w^2)$, where
the relative orbital velocity $V_{orb} \approx 5$~km/s is much less than
the probable stellar wind velocity from the companion,
$V_w \sim 100$~km/s. One
may expect the white dwarf to accrete the wind gas at a rate
$\dot{M_a} = \dot{M_w} (\pi R_c^2 / 4 \pi a^2) \approx 1.6 \cdot 10^{-4}
m_{0.6} v_{100}^{-2}$. Assuming the mass loss rate in the wind of 40~Eridani~C
to be $\dot{M_w} \sim  10^{-15}\, M_{\sun}$/y, we obtain about the
same value of the accretion rate as that found from ISM, $\dot{M_a}
\sim  10^{-19}\, M_{\sun}$/y. Both the circumstellar gas density and
its velocity field around 40~Eridani~B may change in a rather complex way
depending on the companions' winds interaction and orbital phase.

Using the above discussed magnetic field strength and rotational period
we can find whether the accreting gas reaches the stellar surface or not.
If the magnetic field is dipolar with the pole strength $B_p$, the
Alfven radius (radius of stopping of accretion) is
$R_A^{7/2} = B_p^2 R^6 / \dot{M_a} (8 G M)^{1/2}$, where the stellar
radius $R \approx 9 \cdot 10^8$~cm. The co--rotation radius is
$R_c^{1/3} = G M P^2 / 4 \pi^2$, where P is the rotational period.
The accretion is permitted when $R_A < R_c$ or $\dot{M_a} > \dot{M_{cr}}
= 8^{-1/2} (2 \pi)^{7/3} (G M)^{-5/3} P^{-7/3} B_p^2 R^6 \approx
9 \cdot 10^{-20}\, m_{0.6}^{-5/3} p_3^{-7/3} b_5^2\, M_{\sun}$/y,
where $p_3 = P/3$\,hours, $b_5 = B_p/5$\,kG.

We find that 40~Eridani~B
can  accrete the gas both from the ISM and
from the companions onto the surface. At the same time, because we have found
that $\dot{M_a} \,\, \gea \,\, \dot{M_{cr}}$, the accreting gas will
form a circumstellar rotating envelope in the magnetosphere at a distance
$r \sim R_A \sim 4 \cdot 10^{11}$~cm. The circumstellar
gas can  produce the ultraweak absorption spectrum observed.


\begin{acknowledgements}
The work was supported by the RFBR grant N\,98--02--16554.
\end{acknowledgements}

{}

\end{document}